\documentclass[runningheads]{llncs}
\usepackage[T1]{fontenc}
\usepackage{graphicx}
\usepackage{booktabs}
\usepackage[misc]{ifsym}
\newcommand{\corr}{(\Letter)}
\usepackage{tikz}%

\usetikzlibrary{positioning, fit, backgrounds, arrows.meta}
\usepackage{tabularx}
\usepackage{multirow}%
\newcolumntype{L}[1]{>{\hsize=#1\hsize\raggedright\arraybackslash}X}
\newcolumntype{C}[1]{>{\hsize=#1\hsize\centering\arraybackslash}X}
\newcolumntype{R}[1]{>{\hsize=#1\hsize\raggedleft\arraybackslash}X}

\usepackage{amsmath,amssymb,amsfonts}%
\usepackage{mathrsfs}%
\usepackage[title]{appendix}%
\usepackage{xcolor}%
\usepackage{textcomp}%
\usepackage{manyfoot}%
\usepackage{booktabs}%
\usepackage{listings}%
\usepackage{todonotes}%
\usepackage{tikz}
\usepackage{hyperref}
\usepackage[depth=2]{bookmark}
\usepackage{subcaption}

\usepackage{pifont}

\usepackage{color-edits}
\addauthor{gemini}{red}
\addauthor{jakub}{blue}
\addauthor{gustav}{magenta}
\addauthor{oleksii}{green}

\usepackage{soul}
\newcounter{mycomment}

\newcommand{\figpanel}[2]{Fig.~\hyperref[#1]{\getrefnumber{#1}#2}}

\usepackage[normalem]{ulem}

\begin{document}

\title{Relational Task Generation Language:\\ A Declarative Specification Framework for Relational Deep Learning}

\titlerunning{Relational Task Generation Language}

\author{
    Oleksii Kolesnichenko \and Jakub Pele\v{s}ka  \corr \and Gustav \v{S}\'{\i}r 
}
\authorrunning{O. Kolesnichenko et al.}

\institute{Czech Technical University in Prague,\\
Karlovo náměstí 13, Prague, 121 35, Czechia
\email{kolesole@fel.cvut.cz,\{jakub.peleska,gustav.sir\}@cvut.cz}
}

\maketitle

\begin{abstract}
Relational Deep Learning (RDL) has become a powerful paradigm for learning from multi-tabular data. However, manually defining RDL prediction tasks is a laborious process that frequently results in data leakage. To address this issue, we introduce the Relational Task Generation Language (RTGL), an \textit{open-source} declarative language that streamlines RDL task formulation by abstracting away low-level SQL details. We showcase RTGL by reconstructing existing RDL benchmark tasks and uncovering their inconsistencies stemming from hand-written SQL definitions of RDL prediction targets, thereby underscoring the value of a dedicated declarative language. In addition, we demonstrate the practical utility of RTGL by designing various new tasks with diverse forms and target types. Our experiments confirm the robustness and usability of RTGL, as well as its integration with existing RDL frameworks, making it widely accessible to the community.

\keywords{Relational Deep Learning \and Domain-Specific Language \and Predictive Task Generation \and Graph Neural Networks \and Relational Databases }

\end{abstract}

\section{Introduction}
\label{sec:introduction}

Relational databases (RDBs) are ubiquitous for storing and managing structured data across modern enterprises~\cite{codd1990relational}. Yet, leveraging deep learning~\cite{goodfellow2016deep} on this data remains challenging~\cite{borisov2022deep}. While classic machine learning (ML) domains such as computer vision and natural language processing readily represent data as fixed-size numeric tensors~\cite{krizhevsky2012imagenet,he2016deep,vaswani2017attention,devlin2019bert}, relational data, fragmented across multiple interrelated tables of diverse shapes and forms, is fundamentally different. To adapt relational data for classic ML, practitioners have traditionally performed complex joins and aggregations to first denormalize the RDBs~\cite{kramer2001propositionalization} into a suitable format. However, this manual feature engineering is not only time-consuming but also inevitably discards information while disrupting the original RDB structure.

To learn from these multi-table structures without the need for manual feature flattening, Relational Deep Learning (RDL)~\cite{zahradnik2023deep,fey2024position} has recently emerged as a new ML paradigm. Rather than forcing the data into a fixed-form tensor, RDL treats the entire database as a temporal heterogeneous graph~\cite{robinson2024relbench}. In this representation, tables are translated into distinct node types, and the foreign-key relationships between them define the edges. Once this graph is constructed, it can be directly processed by specialized graph neural architectures~\cite{peleska_tabular_2025,chen_relgnn_2025,ranjan_relational_2025} designed to capture both the semantic features of individual rows and the structural topology of the RDB.

While RDL offers an elegant model architecture blueprint, the practical bottleneck has moved to defining the prediction tasks themselves. In relational learning~\cite{Raedt} in particular, training depends on precisely specified prediction problems---what entity to forecast, which label to target, and the exact timestamp of prediction. Deriving such training data demands complex, time-aware SQL~\cite{chamberlin_sequel_1974} queries, where engineers must enforce temporal cutoffs so that the model sees only past data, fully isolated from future outcomes~\cite{robinson2024relbench,peleska_redelex_2025}. This largely manual process is error-prone and often leads to temporal leakage, where information from the future unintentionally contaminates the training features.

Recently, the \textit{Predictive Query Language} (PQL)~\cite{kocijan_predictive_2026} was introduced to simplify task generation in RDL by providing a high-level abstraction specifically tailored for predictive modeling on relational data. However, PQL remains a proprietary, closed-source language, deeply embedded within a specific commercial platform,\footnote{Developed by the team at \href{https://kumo.ai}{Kumo.ai}.} which severely limits its accessibility, transparency, and extensibility for the broader RDL research community. 

To address this limitation, we introduce the Relational Task Generation Language (RTGL), an \textit{open-source} declarative language for generating tasks in RDL, which we make publicly available\footnote{At \url{https://github.com/kolesole/RTGL}; installable as a package from PyPI via \texttt{pip install rtgl}.} to the community. While RTGL largely adopts the syntax and conceptual goals of PQL, it is an independent project designed to integrate with existing open-source RDL frameworks, such as RelBench~\cite{robinson2024relbench} and \textsc{ReDeLEx}~\cite{peleska_redelex_2025}. Beyond the functionality of PQL, RTGL resolves relational paths of arbitrary length, admits explicit path declarations where that resolution is ambiguous, and offers a raw-SQL escape hatch where its grammar falls short (Sec.~\ref{sec:rtgl:paths}).

\begin{figure}[th]
    \centering
    \includegraphics[width=.82\linewidth]{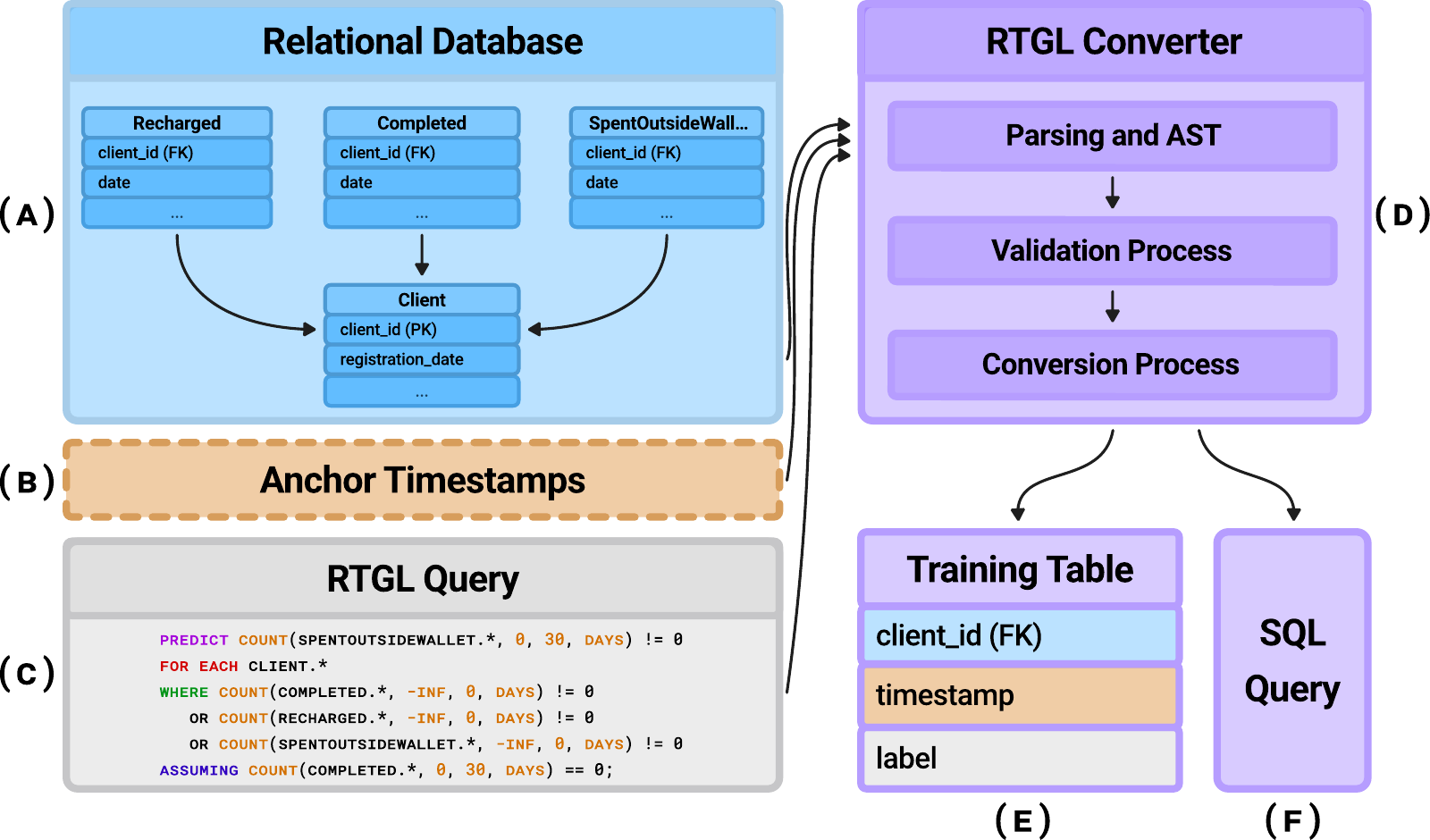}
       \caption{\textbf{RTGL Pipeline.} Schematic overview of the system architecture: the inputs---{(A)} a relational database, {(B)} optional anchor timestamps for temporal tasks, and {(C)} an RTGL query---are processed by {(D)} the three-stage RTGL converter engine, which outputs either {(E)} a materialized training table or {(F)} a raw, time-safe SQL query.}
    \label{fig:pipeline}
\end{figure}

RTGL allows researchers to define complex temporal prediction tasks with only a few lines of declarative code, as depicted in Fig.~\ref{fig:pipeline}. We demonstrate RTGL’s capabilities by reconstructing existing RelBench~\cite{robinson2024relbench} tasks and identifying inconsistencies arising from manually written, complex SQL definitions of RDL prediction targets---highlighting the value of a new dedicated language. We further showcase RTGL by formulating various new tasks within the \textsc{ReDeLEx} framework~\cite{peleska_redelex_2025} with diverse target types, emphasizing its practical utility.

\section{Related Work}
\label{sec:related-work}

While tabular learning methods (e.g., TabNet~\cite{arik2021tabnet}, tabular ResNet~\cite{gorishniy_revisiting_2021}) and in-database machine learning extensions (e.g., BigQuery ML~\cite{bigqueryml}, PostgresML~\cite{postgresml}, SQL4ML~\cite{makrynioti2019sql4ml}) have simplified model deployment, they remain constrained by their requirement for a single, flat input table. To capture the heterogeneity of relational databases, recent advances in RDL have shifted toward integrating tabular representation learning~\cite{lachi_boosting_2025,peleska_tabular_2025} with graph representation learning~\cite{zhou2018graph}, large language models~\cite{wu_large_2025}, and graph transformers~\cite{dwivedi_relational_2026}. This progress is driving the field toward RDL foundation models~\cite{wehrstein_towards_2025,dwivedi_challenges_2025,fey_kumorfm_2025,hudovernik_kumorfm-2_2026,wang_griffin_2025,dwivedi_relational_2026}, supported by task-agnostic pretraining~\cite{peleska_task-agnostic_2026} and synthetic data generation~\cite{jurkovic2025syntherela,kothapalli_plurel_2026}. To support this architectural evolution, standardized benchmarking ecosystems like RelBench~\cite{robinson2024relbench,gu_relbench_2026} and \textsc{ReDeLEx}~\cite{peleska_redelex_2025} have formalized predictive tasks with chronological splits to prevent temporal leakage~\cite{kocijan_predictive_2026} and expanded evaluation across diverse databases to analyze structural impacts.

\section{Relational Deep Learning}
\label{sec:rdl}

Relational Deep Learning is a machine learning paradigm built on interpreting relational databases as heterogeneous graphs. This perspective enables direct application of deep learning techniques---specifically graph neural networks alongside tabular representation learning---to the native structures of RDBs. While the underlying idea was explored in earlier work~\cite{vsir2021deep,Blockeel,aschenbrenner2013deep}, it was recently formalized under the RDL umbrella~\cite{fey2024position} to consolidate these efforts into a cohesive research field, detailed below.

\subsection{Graph Representation}
\label{sec:rdl:graph}
A relational database $\mathcal{R}$ is a finite set of relations $R_1, R_2, \dots, R_n$, each represented as a table $T_R$ whose rows are the tuples $t_i = (a_1, \ldots, a_k)$, that is, the entities of the relation. The attributes of each table fall into four functional groups: a \textit{primary key} $PK_R$, a minimal attribute set uniquely identifying every entity; \textit{foreign keys} $FK_{R_i \to R_j}$, attributes of $R_i$ referencing the primary key of $R_j$ and thereby ensuring referential integrity; \textit{feature attributes}, carrying the characteristic information of the entity; and, optionally, a \textit{timestamp} recording when the row was created.

The \textit{relational entity graph} of $\mathcal{R}$ is then a temporal heterogeneous graph $G = (\mathcal{V}, \mathcal{E}, \phi, \psi, \tau)$. Its nodes $\mathcal{V} = \{v_{i,j} \mid R_i \in \mathcal{R}, t_j \in R_i\}$ are the tuples of every relation, so that $v_{i,j} \simeq t_j \in R_i$, while its edges $\mathcal{E}$ are induced by the primary-key to foreign-key references, an edge arising whenever $t_k[FK_{R_j}] = t_l[PK_{R_j}]$ or conversely. Heterogeneity is captured by three mappings: the node mapping $\phi: \mathcal{V} \to \mathcal{T}^v$ sends each node to its source relation, so that the node types $\mathcal{T}^v$ correspond exactly to the relations of $\mathcal{R}$; the edge mapping $\psi: \mathcal{E} \to \mathcal{T}^e$ sends each edge to the pair of key attributes, or their reverse, that generated it; and the time mapping $\tau: \mathcal{V} \to \mathrm{T}$ assigns each node its timestamp, defaulting to zero where no temporal attribute is available.

\subsection{Neural Models}
\label{sec:rdl:models}

Building upon the graph representation $G$, RDL models typically follow a four-stage pipeline: (i) a \textit{table-level attribute encoder} initializes node embedding matrices $h_v^{(0)}$ by encoding each feature attribute of the node's source relation $\phi(v)$ according to its semantic data type; (ii) a \textit{table-level tabular model} refines these into $h_v^{(l)}$, optionally compressing the per-attribute matrix into a single vector; (iii) a \textit{graph neural model} applies message passing over the edge types $\mathcal{T}^e$, using standard graph neural networks for vector embeddings or custom schemes for attribute matrices; and (iv) a \textit{task-specific model head}, typically a simple multi-layer perceptron, maps the final node embeddings to predictions.

\subsection{Predictive Tasks}
\label{sec:rdl:tasks}

\begin{figure}[t]
  \centering
  \includegraphics[width=.52\linewidth]{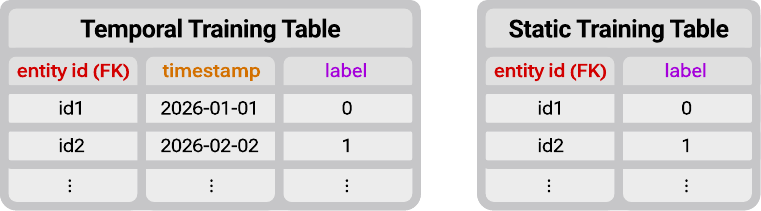}
  \caption{\textbf{Training Tables.} Illustration of training tables for temporal and static prediction tasks. Both table types contain an \texttt{entity id} foreign key that references the target table (e.g., a client id), and a \texttt{label} column specifying the prediction target (e.g., number of client orders). The temporal training table additionally includes a \texttt{timestamp} column, indicating when the prediction is made.}
  \label{fig:training_tables}
\end{figure}

Predictive tasks in RDL are formalized by constructing a dedicated \textit{task table} $T_{\text{task}}$ that extends the existing database $\mathcal{R}$~\cite{robinson2024relbench,kocijan_predictive_2026}. Its tuples are the individual prediction instances $(v, y, t)$, where $v \in \mathcal{V}$ is the target entity identified via a foreign-key reference $T_{\text{task}}[FK]$, $y \in \mathcal{Y}$ is the ground-truth label to be predicted (e.g., $y \in \mathbb{R}$ for regression), and $t \in \mathrm{T}$ is the anchor time at which the prediction is made. The objective of an RDL model $f_\theta$ is then to predict $\hat{y} = f_\theta(G_{v, \leq t}) \approx y$, where $G_{v, \leq t}$ is the relational subgraph forming the computational context for $v$, restricted to information available at or before $t$. Depending on the role of $t$, tasks are categorized as static or temporal, with the training tables of both illustrated in Fig.~\ref{fig:training_tables}.

\paragraph{Static Tasks.} Static tasks predict missing values or properties within a database viewed as a single fixed snapshot in time, reducing instances to pairs $(v, y)$ and the objective to $\hat{y} = f_\theta(G_v)$ over the entire available graph $G$, with no risk of temporal leakage. The target $y$ here forms a feature column that already structurally exists within the database but is masked out.

\paragraph{Temporal Tasks.} Conversely, temporal (or ``forecasting'') tasks predict events that have not yet occurred, so the anchor time $t$ becomes crucial: the label $y$ is an aggregation or occurrence in a future window $(t, t + \Delta w]$, and the label set is $Y = \{Q_{\Delta w}(t) \mid t \in \mathrm{T}_{\text{task}}\} \subseteq \mathcal{Y}$ for a future-pointing query $Q$ of fixed window size $\Delta w$ over the anchor timestamps $\mathrm{T}_{\text{task}}$. Preventing temporal leakage requires conditioning only on past data, that is, masking the graph to $G_{\leq t} = (\mathcal{V}_{\leq t}, \mathcal{E}_{\leq t})$ with $\mathcal{V}_{\leq t} = \{u \in \mathcal{V} \mid \tau(u) \le t\}$.

\section{Relational Task Generation Language}
\label{sec:RTGL}

To automatically construct the training tables~(Fig.~\ref{fig:training_tables}) needed for RDL~(Sec.~\ref{sec:rdl:tasks}), we developed RTGL, modeled on the features of the proprietary PQL~\cite{kocijan_predictive_2026}. The language satisfies three criteria: it is \textit{declarative}, enabling users to state what to predict rather than how to obtain it; it enforces \textit{temporal safety} by clearly distinguishing historical from predictive time windows to avoid data leakage~\cite{hadsell2020embracing}; and it is broadly applicable across \textit{different task types}, including classification, regression, and link prediction. While RTGL adopts PQL's overall structure, syntactic patterns, and keywords,\footnote{For a more detailed account of PQL, we refer the reader to Kumo.ai’s official documentation at \url{https://kumo.ai/docs/predictive-query/} or the associated publication~\cite{kocijan_predictive_2026}.} it is an independent, \textit{open-source} project.

\begin{figure}[t]
    \centering
    \includegraphics[width=.80\linewidth]{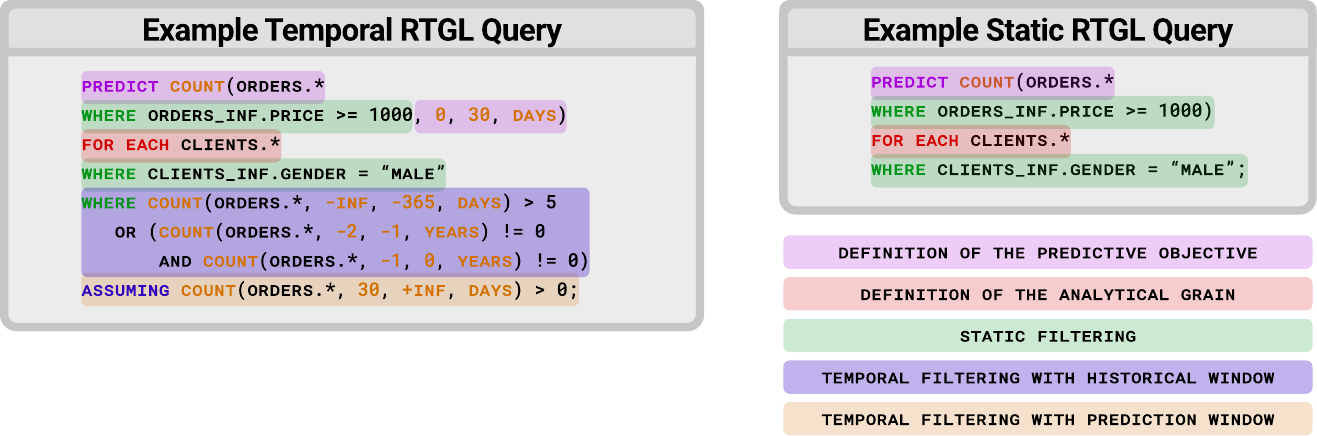}
    \caption{\textbf{RTGL Query Examples.} In both queries, the goal is to predict the number of orders with a price of at least 1000 placed by male clients. The static query (right) evaluates this condition over the entire database, whereas the temporal query (left) adds constraints that restrict the aggregation to a 30-day prediction window and filter clients through historical observation windows, all relative to the anchor timestamp.}
    \label{fig:rtgl_example_highlights}
\end{figure}

\subsection{Language Syntax}

Using an SQL-like syntax~\cite{chamberlin_sequel_1974}, RTGL expresses predictive RDL tasks by specifying what to predict, for which entity, and within what (temporally safe) future and historical time context. 

The \texttt{PREDICT} clause serves as the entry point for defining the predictive objective of the task, dictating the target label to be computed for each prediction instance (Sec.~\ref{sec:rdl:tasks}), as illustrated in the query of Fig.~\ref{fig:rtgl_example_highlights}, where the objective is to predict the \texttt{COUNT} of \texttt{orders.*}. This clause supports mathematical aggregations, logical conditions, and, in static queries, direct references to raw column values. For temporal queries (Fig.~\ref{fig:rtgl_example_highlights}), all aggregations in this clause must reference a defined future time window to prevent leakage.

The \texttt{FOR EACH} clause defines the primary entity table (Sec.~\ref{sec:rdl:tasks}) and the specific identifier column that acts as the focal point for the prediction. By binding the extraction logic to a specific node type in the heterogeneous graph (Sec.~\ref{sec:rdl:graph}), this clause ensures that the resulting task table accurately maps predictions back to the correct entities via their primary keys. In Fig.~\ref{fig:rtgl_example_highlights}, \texttt{FOR EACH clients} binds the predictions to individual clients, establishing the analytical grain of the predictive task. In a temporal query, the entity table's time column additionally acts as a mask---\textit{temporal masking}---emitting an entity only at anchor timestamps at or after its own time value, so no prediction is made for a client before they registered.

Aggregation operators, such as \texttt{SUM}, \texttt{COUNT}, and \texttt{AVG}, compress future or historical records into single scalar values. As summarized in Table~\ref{tab:rtgl_aggregations}, RTGL supports a range of mathematical and logical operators tailored to specific data types. For temporal tasks, these aggregations explicitly require defined start and end time shifts, forcing the user to map out the chronological bounds of the operation. Used directly, an aggregation yields a regression target; combined with a comparison operator, it yields a binary classification target. A window \texttt{(start, end, UNIT)} denotes the half-open interval $(t+\textit{start},\, t+\textit{end}]$ relative to the anchor timestamp $t$, with bounds signed by clause: forward ($\geq 0$) in \texttt{PREDICT} and \texttt{ASSUMING}, backward ($\leq 0$) in the temporal \texttt{WHERE}. Both are mandatory, so a window left open at one end is inexpressible.

The \texttt{ASSUMING} clause governs the future prediction window. This clause applies filtering logic directed toward future events relative to the anchor timestamp. For instance, the clause \texttt{ASSUMING COUNT(orders.*, 30, +INF, DAYS) > 0} in the temporal query of Fig.~\ref{fig:rtgl_example_highlights} restricts the result to entities that still have at least one order occurring after the 30-day prediction window.

The \texttt{WHERE} clause governs the historical observation window. By looking only backward from the anchor timestamp, it bounds the historical context needed for prediction, allowing the framework to select training instances according to prior behaviors or conditions. In Fig.~\ref{fig:rtgl_example_highlights}, for instance, the clause \texttt{WHERE COUNT(orders.*, -1, 0, YEARS) != 0} limits the training data to include only clients who have placed at least one order in the year preceding the anchor timestamp. Additionally, \texttt{WHERE} can be used for static filtering, without temporal constraints, in which case its behavior coincides with that of standard SQL~\cite{chamberlin_sequel_1974}.

Finally, the \texttt{LIST\_DISTINCT} operator is used for tasks requiring set-based predictions, such as multilabel classification or link prediction. Unlike standard aggregations that return a scalar value, this operator returns a set of distinct categorical values or foreign keys over a specified temporal window. This feature is important for scenarios where the objective is to predict multiple future interactions, such as recommending a list of items a user might purchase.

\begin{table}[t]
    \centering
    \begin{tabular}{p{1.8cm}p{4.5cm}l}
    \toprule
    \textbf{Type} & \textbf{Aggregation} & \textbf{Condition} \\ 
    \midrule
    Numerical & \texttt{AVG}, \texttt{COUNT}, \texttt{COUNT\_DISTINCT}, & \texttt{!=}, \texttt{<}, \texttt{<=}, \\
              & \texttt{FIRST}, \texttt{LAST}, \texttt{MAX}, & \texttt{==}, \texttt{>}, \texttt{>=} \\ 
              & \texttt{MIN}, \texttt{SUM} & \\ 
    \addlinespace
    \addlinespace
    String & \texttt{FIRST}, \texttt{LAST} & \texttt{NOT LIKE}, \texttt{NOT CONTAINS}, \texttt{ENDS WITH}, \\
    & & \texttt{STARTS WITH}, \texttt{LIKE}, \texttt{CONTAINS}, \texttt{=} \\
    \addlinespace
    Nullable & \texttt{AVG}, \texttt{COUNT}, \texttt{COUNT\_DISTINCT}, & \texttt{IS NULL}, \texttt{IS NOT NULL} \\
         & \texttt{FIRST}, \texttt{LAST}, \texttt{MAX}, & \\ 
         & \texttt{MIN}, \texttt{SUM}, \texttt{LIST\_DISTINCT} & \\ 
    \bottomrule
    \hspace{0.5cm}
\end{tabular}
    \caption{\textbf{Aggregation and Condition Functions in RTGL.} The exhaustive list of aggregations and conditions available in RTGL for each data type, mapping closely to the foundational capabilities outlined in PQL.}
    \label{tab:rtgl_aggregations}
\end{table}

\subsection{Relational Path Resolution}
\label{sec:rtgl:paths}
\paragraph{Automatic Multi-Hop Resolution.} RTGL queries need not state \textit{how} two tables are connected: the engine derives the connection from the foreign keys declared in the database schema, searching for the \textit{shortest path of foreign keys} between the aggregated table and the \texttt{FOR EACH} entity. Such a chain may be of any length and may traverse foreign keys in either direction, which keeps a clause such as \texttt{COUNT(posts.*, 0, 30, DAYS)} valid however many bridge tables separate \texttt{posts} from the predicted entity. The resolution is deliberately conservative: a unique shortest path is used silently; if competing paths \textit{differ} in length, the shortest is chosen and reported as a warning; and if several shortest paths of \textit{equal} length exist, the query is rejected rather than resolved arbitrarily. Temporal queries further admit only paths through at least one table with a time column other than the entity table, so the same pair of tables may resolve differently in static and temporal queries.

\paragraph{Common Path Expressions.} Where the schema itself is ambiguous, \textit{common path expressions} (CPEs) name an explicit join path that is then referenced like an ordinary table. CPEs are declared in an optional \texttt{WITH} clause preceding all others. Each path is a chain of \texttt{->} hops between nodes, where every node is a table together with the key on which it is \textit{entered} and an optional \texttt{:<right\_key>} suffix naming the key through which it is \textit{left}. A reference to the alias resolves against the \textit{source} table of the path, never its destination. For instance, the path \texttt{circuits.\allowbreak circuitId -> races.\allowbreak circuitId:\allowbreak raceId -> results.\allowbreak raceId:\allowbreak driverId -> drivers.\allowbreak driverId} is precisely what the \textit{driver-circuit-compete} task of \textit{rel-f1} requires: three routes of identical length connect \texttt{circuits} to \texttt{drivers} through \texttt{results}, \texttt{standings}, or \texttt{qualifying}, only the first of which records that a driver took part in a race.

\paragraph{SQL Injections.} CPEs presuppose a declared foreign key and a join expressible in the \texttt{table.key -> table.key} shape of the grammar. Where either assumption fails, an \textit{SQL injection} supplies raw SQL wherever a table is expected, and the engine treats its result as a \textit{virtual table}. Syntactically, an injection is a body enclosed in square brackets, followed by positional brace-delimited slots declaring the name under which the result is exposed, its primary key, its outgoing foreign keys with their parent tables, the incoming references from other tables, and its time column. Each injection compiles into a further CTE whose descriptor makes it indistinguishable from a real table for path resolution and temporal windowing, though its body is deliberately left unvalidated. Injections thereby cover the cases in which a query cannot even be \textit{stated}: an undeclared relationship, an SQL function the language does not expose, and a per-child aggregate requiring re-aggregation at the parent level.

\section{Internal Design and Implementation}
\label{sec:implementation}

The internal implementation of RTGL translates the declarative syntax (Sec.~\ref{sec:RTGL}) into executable SQL while guaranteeing semantic and chronological correctness. As depicted in the RTGL pipeline (Fig.~\ref{fig:pipeline}), the converter engine (\figpanel{fig:pipeline}{D}) operates on a raw RTGL query (\figpanel{fig:pipeline}{C}), a relational database schema (\figpanel{fig:pipeline}{A}), and optional anchor timestamps (\figpanel{fig:pipeline}{B}). To keep the design modular, the engine is conceptually divided into three sequential steps: parsing and Abstract Syntax Tree (AST) generation, semantic validation, and query conversion.

\subsection{Parsing and AST Generation}
The first step transforms the raw query string (\figpanel{fig:pipeline}C) into a structured format. This phase is implemented with the ANTLR4~\cite{parr2013antlr4} parser generator, chosen for its adaptive LL(*) parsing algorithm and its ability to automatically generate a Concrete Syntax Tree (CST) based on the formal grammar. However, the raw CST contains redundant syntactic artifacts. To extract meaningful semantic logic, the engine uses the Visitor pattern. This process selectively explores the tree and dynamically constructs an AST, standardizing the representation into a nested dictionary. This structure effectively isolates the raw text parsing from the deeper logical operations while preserving positional metadata (line and column numbers) necessary for precise error reporting.

\subsection{Semantic Validation}
While ANTLR4 enforces syntactic correctness, a grammatically valid query is not necessarily semantically sound or executable within the context of the given database. The generated AST is thus passed to the validation process, which ensures logical correctness against the provided relational schema (\figpanel{fig:pipeline}{A}). The validation module verifies the existence of all referenced tables and columns and ensures that primary- and foreign-key constraints are valid. Notably, the process performs strict type checking, verifying that the data types of the aggregations match the conditional operators applied to them. Furthermore, temporal validation enforces chronological consistency: observation windows must use non-positive time shifts (the past), whereas prediction windows must use non-negative time shifts (the future). The same stage performs relational path resolution (Sec.~\ref{sec:rtgl:paths}), searching the foreign-key graph for the shortest connection between each aggregated table and the entity table, and rejecting the query where several candidates of equal length remain or none exists; injection bodies are the only exception, as they are opaque to the engine by design. Validation also deliberately does not \textit{fail fast}: rather than halting at the first detected error, it accumulates all syntactic and semantic violations across the entire AST to provide a comprehensive error report before any data extraction occurs.

\subsection{Query Conversion}
Once the query passes all semantic checks, the AST is processed by the conversion module to materialize the executable logic. Instead of generating a single, massive SQL statement, the converter dynamically constructs isolated relational algebra subtables for distinct logical expressions by recursively unwrapping the AST dictionary. These subtables are then combined using formal relational operations; for instance, Boolean \texttt{AND} conditions translate into SQL \texttt{INTERSECT} operations, while \texttt{OR} conditions translate into \texttt{UNION} operations. For temporal tasks, the conversion engine uses Common Table Expressions (CTEs)~\cite{melton2002advanced} to place the anchor timestamps (\figpanel{fig:pipeline}{B}) at the top of the SQL query. All subsequent intermediate subtables and historical observation windows are computed relative to these anchor timestamps, preventing temporal leakage at the execution level. Common path expressions and SQL injections are materialized by the same mechanism, each emitted as a further CTE registered under an internal table descriptor, which lets the remainder of the pipeline treat both as ordinary tables. The final output is returned either as a materialized training table (\figpanel{fig:pipeline}{E}) or as a raw, time-safe SQL query (\figpanel{fig:pipeline}{F}), the former obtained by executing the generated statement on an embedded DuckDB engine.\footnote{\url{https://duckdb.org}}

\section{Validation and Experiments}
\label{sec:experiments}

The primary objective of our experiments is to demonstrate that RDL task generation via RTGL is simple, correct, and practically viable. To validate the framework, we (i) verify RTGL empirically against the established RelBench benchmark~\cite{robinson2024relbench}, (ii) demonstrate its extensibility on the \textsc{ReDeLEx} framework~\cite{peleska_redelex_2025}, and (iii) confirm the structural integrity of the generated tasks via end-to-end model training. All tasks and the experiment code are available on GitHub.\footnote{\url{https://github.com/kolesole/rtgl-tasks}}

\begin{figure}[t]
    \centering
    \begin{subfigure}[t]{.31\textwidth}
        \centering
        \includegraphics[width=\linewidth]{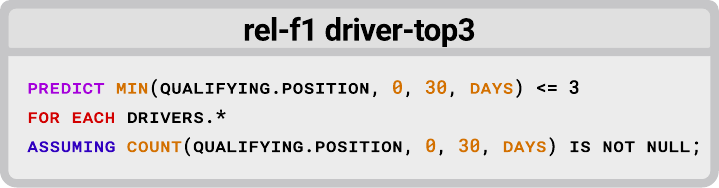}
        \caption{\textbf{driver-top3.}}
        \label{fig:relf1_drivertop3}
    \end{subfigure}%
    \hfill
    \begin{subfigure}[t]{.31\textwidth}
        \centering
        \includegraphics[width=\linewidth]{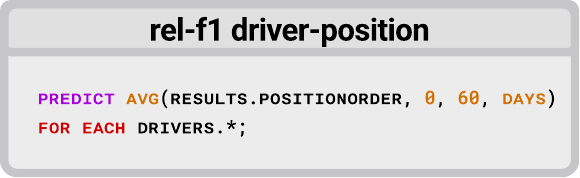}
        \caption{\textbf{driver-position.}}
        \label{fig:relf1_driverposition}
    \end{subfigure}%
    \hfill
    \begin{subfigure}[t]{.31\textwidth}
        \centering
        \includegraphics[width=\linewidth]{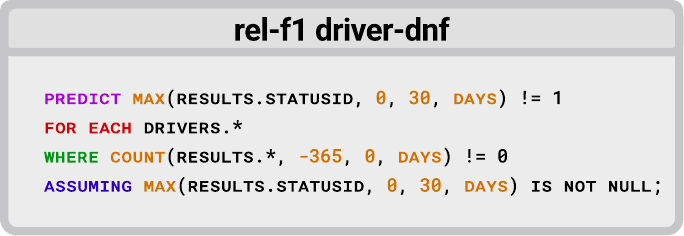}
        \caption{\textbf{driver-dnf.}}
        \label{fig:relf1_driverdnf}
    \end{subfigure}
    \caption{Recreated rel-f1 dataset tasks from the RelBench~\cite{gu_relbench_2026} benchmark: (a) binary classification of whether a driver will qualify in the top 3 for a race in the next month; (b) regression of a driver's average finishing position over the next two months; and (c) binary classification of whether a driver will fail to finish a race (DNF) in the next month.}
    \label{fig:relf1_tasks}
\end{figure}

\subsection{Recreated Tasks}
\begin{figure}[t]
    \centering
    \begin{subfigure}[t]{.31\textwidth}
        \centering
        \includegraphics[width=\linewidth]{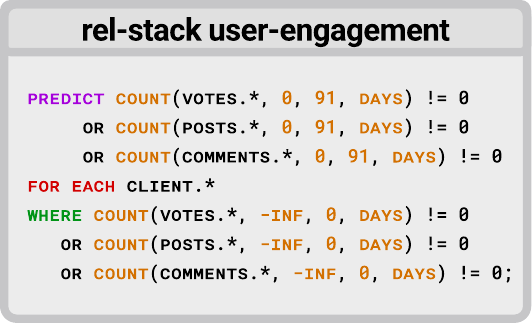}
        \caption{\textbf{user-engagement.}}
        \label{fig:relstack_usereng}
    \end{subfigure}%
    \hfill
    \begin{subfigure}[t]{.31\textwidth}
        \centering
        \includegraphics[width=\linewidth]{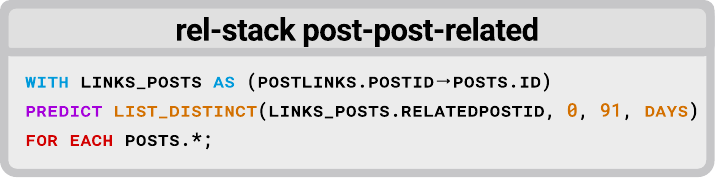}
        \caption{\textbf{post-post-related.}}
        \label{fig:relstack_postrelated}
    \end{subfigure}%
    \hfill
    \begin{subfigure}[t]{.31\textwidth}
        \centering
        \includegraphics[width=\linewidth]{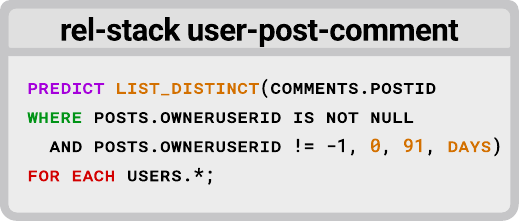}
        \caption{\textbf{user-post-comment.}}
        \label{fig:relstack_usercomment}
    \end{subfigure}
    \caption{Recreated rel-stack dataset tasks from the RelBench benchmark: (a) binary classification of whether a user will make any votes, posts, or comments in the next three months; (b) link prediction of the existing posts that users will link a given post to, and (c) of the existing posts a user will comment on, both over the next two months.}
    \label{fig:relstack_tasks}
\end{figure}

To establish an empirical baseline and verify the correctness of RTGL, we regenerated several predefined tasks from the RelBench benchmark across the \textit{rel-f1} (Fig.~\ref{fig:relf1_tasks}) and \textit{rel-stack} (Fig.~\ref{fig:relstack_tasks}) datasets. The majority of standard predictive tasks were successfully recreated using simple declarative statements. 

To empirically demonstrate the consistency of the generated relational tables, we joined them with the RelBench reference tables to identify any discrepancies. This direct comparison confirms equivalence between the RTGL outputs and the established baselines for correctly formulated tasks. Tasks whose targets are separated from the predicted entity by several relational hops are covered natively (Sec.~\ref{sec:rtgl:paths}). The two residual cases are handled with explicit path declarations (CPEs) rather than automatic search: \textit{driver-circuit-compete}, whose three equally short routes the engine rightly refuses to disambiguate, and the self-referential \textit{post-post-related}. RTGL thus keeps every table referenced in a task definition explicit, without paying for that transparency in expressiveness.

\subsection{Found Inconsistencies}
Redefining these tasks in RTGL had an unexpected outcome: it revealed that the official RelBench benchmark\footnote{\url{https://relbench.stanford.edu}} contains temporal inconsistencies in its manually written SQL task definitions. By comparing the tables, we found cases in which the original RelBench tables incorrectly contained more rows than the temporally sound RTGL tables.

For instance, in the \textit{driver-dnf}~(Fig.~\ref{fig:relf1_driverdnf}) binary classification task, analysis of the original SQL query revealed an error in its temporal logic. The query filtered for active drivers but failed to apply an upper-bound constraint relative to the anchor timestamp. Consequently, the query looked into the future, filtering the main table based on driver activity that had not yet occurred at prediction time, and occasionally included drivers that did not yet exist at that time.

A similar mismatch occurred during the validation of the \textit{user-engagement} and \textit{user-post-comment} tasks. In these tasks, the original RelBench SQL failed to filter users by their creation date. As a result, the baseline tables included records for users who had not yet registered at the given prediction timestamps. This demonstrates that manually writing temporal SQL queries is an error-prone process; RTGL's mandatory bounds rule such errors out by construction. Both defects stem from an omitted bound, which RTGL does not allow: a window is a pair, so a filter constraining the past while leaving the future open cannot be written, and temporal masking drops entities that do not yet exist at an anchor timestamp.

\begin{figure}[t]
\centering
\begin{subfigure}[t]{0.49\textwidth}
    \centering
    \includegraphics[width=\linewidth]{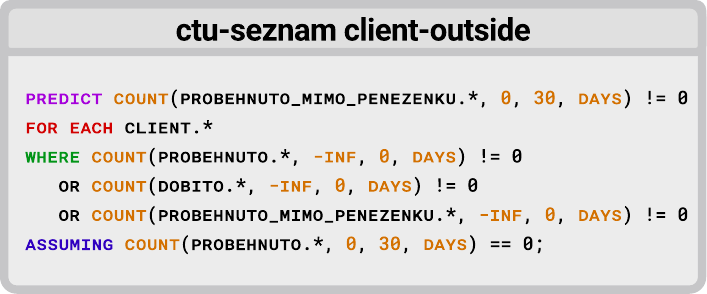}
    \caption{\textbf{client-outside.}}  
    \label{fig:ctuseznam_clientoutside}
\end{subfigure}\hfill
\begin{subfigure}[t]{0.49\textwidth}
    \centering
    \includegraphics[width=\linewidth]{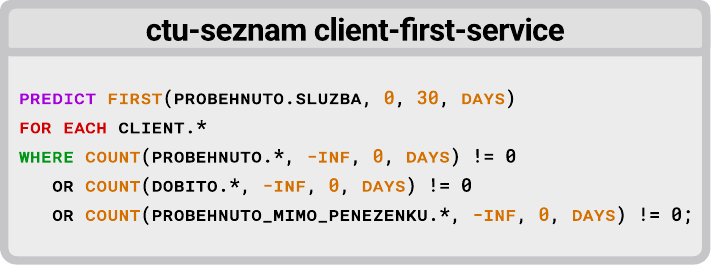}
    \caption{\textbf{client-first-service.}}
    \label{fig:ctuseznam_clientfirstservice}
\end{subfigure}

\vspace{0.4em}

\begin{subfigure}[t]{0.49\textwidth}
    \centering
    \includegraphics[width=\linewidth]{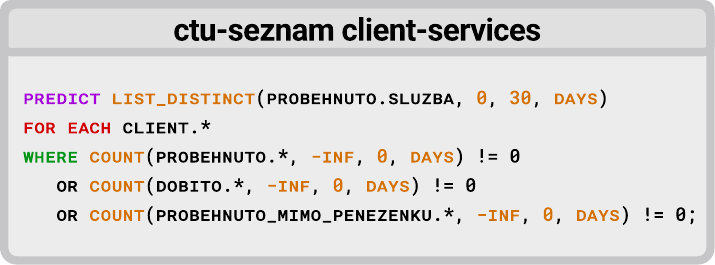}
    \caption{\textbf{client-services.}}
    \label{fig:ctuseznam_clientservices}
\end{subfigure}\hfill
\begin{subfigure}[t]{0.49\textwidth}
    \centering
    \includegraphics[width=\linewidth]{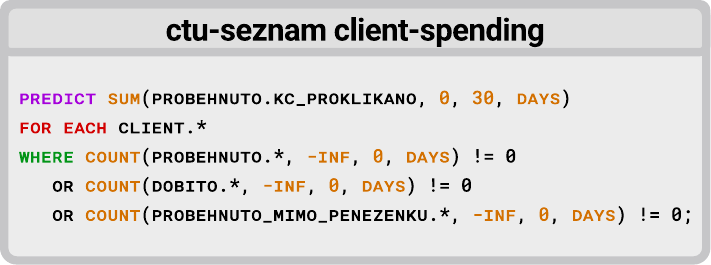}
    \caption{\textbf{client-spending.}}
    \label{fig:ctuseznam_clientspending}
\end{subfigure}

\caption{Novel tasks on the ctu-seznam dataset from the \textsc{ReDeLEx}~\cite{peleska_redelex_2025} framework: (a) binary classification of whether a client will spend money outside the wallet in the next 30 days; (b) multiclass classification of the first service a client will use in the next 30 days; (c) multilabel classification of the services a client will use in the next 30 days; and (d) regression of a client's spending amount over the next 30 days.}
\label{fig:ctuseznam_tasks}
\end{figure}

\subsection{New Tasks}
Following the verification of the framework against established benchmarks, we demonstrate the practical extensibility of RTGL by introducing novel predictive tasks, as shown in Fig.~\ref{fig:ctuseznam_tasks}. These new tasks were formulated for \textit{ctu-seznam}, a complex, enterprise-scale, real-world relational database from the \textsc{ReDeLEx} framework~\cite{peleska_redelex_2025}.\footnote{Derived from the Czech web portal \href{https://www.seznam.cz}{seznam.cz}.} Using RTGL, we defined tasks of four types: binary classification, multiclass classification, multilabel classification, and regression. This shows that RTGL generalizes beyond curated academic benchmarks.

\begin{table}[th]
    \centering
    \begin{tabular}{p{2cm}p{3cm}p{1cm}cc}
\toprule
\multirow{2}{*}{\textbf{Dataset}} & \multirow{2}{*}{\textbf{Task}} & \multirow{2}{*}{\textbf{Split}} & \multicolumn{2}{c}{\textbf{RDL Model}}  \\ 
\cmidrule(lr){4-5}

& & & \textbf{SAGE} & \textbf{HGT} \\
\midrule
\midrule

\multicolumn{5}{c}{{Binary Classification (Accuracy $\uparrow$)}} \\
\midrule
\multirow{2}{*}{rel-f1} & \multirow{2}{*}{driver-dnf} & val & $.8026 \pm .0052$ & $.7996 \pm .0089$ \\
& & test & {.7219} $\pm .0123$ & $.6932 \pm .0138$ \\
\midrule
\multirow{2}{*}{ctu-seznam} & \multirow{2}{*}{client-outside} & val & $.9764 \pm .0016$ & $.9868 \pm 0$ \\
& & test & $.9768 \pm .0016$ & {.9876} $\pm 0$ \\
\midrule
\midrule

\multicolumn{5}{c}{{Multiclass Classification (Accuracy $\uparrow$)}} \\
\midrule
\multirow{2}{*}{ctu-seznam} & \multirow{2}{*}{client-first-service} & val & $.4417 \pm .0058$ & $.4364 \pm .0036$ \\
&  & test & {.4677} $\pm .0056$ & $.4585 \pm .0005$ \\
\midrule
\midrule

\multicolumn{5}{c}{{Regression (MAE $\downarrow$)}} \\
\midrule
\multirow{2}{*}{rel-f1} & \multirow{2}{*}{driver-position} & val & $2.9798 \pm .0803$ & $3.8292 \pm .0154$ \\
& & test & {3.9138} $\pm .0819$ & $4.2645 \pm .0597$ \\
\midrule
\multirow{2}{*}{ctu-seznam} & \multirow{2}{*}{client-spending} & val & $8826.14 \pm 1.11$ & $8839.10 \pm 3.06$ \\
& & test & {9306.87} $\pm 1.54$ & $9321.03 \pm 2.88$ \\
\midrule
\midrule

\multicolumn{5}{c}{{Multilabel Classification (AUPRC $\uparrow$)}} \\
\midrule
\multirow{2}{*}{ctu-seznam} & \multirow{2}{*}{client-services} & val & $.6217 \pm .0017$ & $.5929 \pm .0074$ \\
 &  & test & {.6274} $\pm .0017$ & $.5987 \pm .0078$ \\
\bottomrule
\hspace{0.5cm}
\end{tabular}
    \caption{\textbf{RDL model performance on a selection of RTGL-defined tasks.} Metrics vary depending on the objective: Accuracy for binary and multiclass classification, Mean Absolute Error (MAE) for regression, and Area Under the Precision-Recall Curve (AUPRC) for multilabel classification.}
    \label{tab:results}
\end{table}

\subsection{Results}
To confirm the structural integrity and practical utility of the generated tasks, we selected several redefined tasks from RelBench, alongside the newly defined \textsc{ReDeLEx} tasks, and trained predictive RDL models (Sec.~\ref{sec:rdl:models}) on them. We evaluated two structurally different neural architectures configured with comparable complexity: a GraphSAGE~\cite{hamilton2017inductive} baseline with a mean aggregator and a Heterogeneous Graph Transformer (HGT)~\cite{Hu2020} with two attention heads, representing a more advanced RDL approach. Both models used 128 hidden channels, after which node embeddings were passed through a shared MLP head with 256 hidden units, ReLU nonlinearity, layer normalization, and dropout. All experiments used the AdamW optimizer with a learning rate of 0.0005 and a batch size of 256, training for up to 30 epochs with an early-stopping patience of 10. Neighborhood sizes per layer were 15, 10, and 5 for the new \textsc{ReDeLEx} tasks, whereas a constant size of 5 was applied to all layers for RelBench tasks, to speed up training.

Table~\ref{tab:results} reports the mean and standard deviation of the validation and test metrics. Both models showed strong predictive capabilities and consistent improvement during training across all task types. This successful convergence is consistent with the task tables generated by the RTGL engine being structurally sound and temporally valid. Notably, the simpler GraphSAGE baseline outperformed HGT on five of the six tasks, in line with recent evidence in relational deep learning~\cite{dwivedi_relational_2026}.

\section{Conclusion}
\label{sec:conclusion}

This work addresses the challenge of predictive task generation in relational deep learning. We showed that manual creation of temporal predictive tasks is a laborious process that frequently results in data leakage. We presented RTGL, an open-source declarative language that simplifies predictive task formulation by abstracting away low-level SQL logic. Our experiments demonstrated that recreating standard benchmark tasks with RTGL uncovered temporal inconsistencies in their official manual definitions. Furthermore, RTGL generated valid training tables for novel tasks, enabling the successful training of downstream RDL models. Beyond reproducing the capabilities of PQL, the language resolves relational paths of arbitrary length automatically, admits explicit path declarations where the schema is genuinely ambiguous, and accepts raw SQL as a virtual table where neither schema nor grammar suffices.

While RTGL provides a solid foundation, several limitations remain to be addressed in future work. Most notably, exactly one aggregation is permitted per clause, which precludes re-aggregating a per-child aggregate at the parent level without an injection. Additionally, performance optimization is necessary to accelerate data processing for massive enterprise datasets.

\begin{credits}
\subsubsection*{Acknowledgments.}
This work has received funding from the Czech Science Foundation grant No. 26-22501S. The authors acknowledge the support of the OP VVV funded project CZ.02.1.01/0.0/0.0/16\_019/0000765 ``Research Center for Informatics''.

\end{credits}

\bibliographystyle{splncs04}
\bibliography{references}

\end{document}